\documentstyle[12pt,aasms4]{article}
\tighten
\eqsecnum

\begin{document}

\font\twelvei = cmmi10 scaled\magstep1 
       \font\teni = cmmi10 \font\seveni = cmmi7
\font\mbf = cmmib10 scaled\magstep1
       \font\mbfs = cmmib10 \font\mbfss = cmmib10 scaled 833
\font\msybf = cmbsy10 scaled\magstep1
       \font\msybfs = cmbsy10 \font\msybfss = cmbsy10 scaled 833
\textfont1 = \twelvei
       \scriptfont1 = \twelvei \scriptscriptfont1 = \teni
       \def\mit{\fam1 }
\textfont9 = \mbf
       \scriptfont9 = \mbfs \scriptscriptfont9 = \mbfss
       \def\bmit{\fam9 }
\textfont10 = \msybf
       \scriptfont10 = \msybfs \scriptscriptfont10 = \msybfss
       \def\bmsy{\fam10 }

\def\etal{{\it et al.~}}
\def\eg{{\it e.g.,~}}
\def\ie{{\it i.e.,~}}
\def\lsim{\raise0.3ex\hbox{$<$}\kern-0.75em{\lower0.65ex\hbox{$\sim$}}} 
\def\gsim{\raise0.3ex\hbox{$>$}\kern-0.75em{\lower0.65ex\hbox{$\sim$}}} 
\def\kms{~{\rm km~s^{-1}}}
\def\cm3{~{\rm cm^{-3}}}
\def\yr{~{\rm yr}}
\def\Msun{~{\rm M}_{\sun}}

\def\cf{{\it cf.~}}
\def\cc{CCs~}
\def\coll{ \tau_{coll}}
\def\rad{ \tau_{rad}}
\def\tcr{ \tau_{cr}}
\def\tcs{ \tau_{csc}}
\def\rmm{ \rho_{max}}
\def\roi{ \rho_i}
\def\ycm{ $Y_{cm}$~}
\def\pma{ \p_{max}}
\def\xcoor{{\it X}-coordinate~}
\def\ycoor{{\it Y}-coordinate~}
\def\zcoor{{\it Z}-coordinate~}
\def\ltsima{$\; \buildrel < \over \sim \;$}
\def\simlt{\lower.5ex\hbox{\ltsima}}
\def\gtsima{$\; \buildrel > \over \sim \;$}
\def\simgt{\lower.5ex\hbox{\gtsima}}

\title{Magnetohydrodynamics of Cloud Collisions  \\
in a Multi-phase Interstellar Medium\altaffilmark{4}}

\author{Francesco Miniati \altaffilmark{1},
        Dongsu Ryu \altaffilmark{2},
        Andrea Ferrara \altaffilmark{3}
and     T. W. Jones\altaffilmark{1}}

\altaffiltext{1}{School of Physics and Astronomy, 116 Church Street, SE, 
University of Minnesota, Minneapolis, MN 55455;
\\min@msi.umn.edu, twj@astro.spa.umn.edu}
\altaffiltext{2}{Department of Astronomy \& Space Science, Chungnam National
University, Daejeon 305-764, Korea;
\\ryu@canopus.chungnam.ac.kr}
\altaffiltext{3}{Osservatorio Astrofisico di Arcetri, Largo Enrico Fermi 5, 
I-50125 Firenze, Italy; ferrara@arcetri.astro.it}
\altaffiltext{4}{Submitted to {\it The Astrophysical Journal}}

\begin{abstract}
We extend previous studies of the physics of interstellar cloud 
collisions  by beginning
investigation of the role of magnetic fields 
through 2D magnetohydrodynamical (MHD) numerical simulations. 
In particular, we study head-on collisions between
equal mass, mildly supersonic diffuse clouds similar to our
previous study. Here we include a moderate 
magnetic field, corresponding to $\beta=p_g/p_b=4$,
and two limiting field geometries, with the field
lines parallel (aligned) and perpendicular (transverse)
to the colliding cloud motion. 
We explore both adiabatic and radiative ($\eta=\rad/\coll\simeq 0.38$) cases,
and we simulate collisions between 
clouds evolved through prior motion in the intercloud medium.
Then, in addition to the collision of evolved, 
identical clouds (symmetric cases), 
we also study collisions of initially identical
clouds but with different evolutionary ages (asymmetric cases).

Depending on their geometry, magnetic fields can  significantly alter
the outcome of the collisions compared to the hydrodynamic (HD) case. 
In the {\it (i) aligned} case, adiabatic collisions, like their HD 
counterparts, are very disruptive, independently of the cloud symmetry. 
However, when radiative processes are taken into account, partial coalescence 
takes place even in the asymmetric case, unlike the HD calculations. 
In the {\it (ii) transverse} case, the effects of the magnetic field 
are even more dramatic, with remarkable differences between
unevolved and evolved clouds.
Collisions between (initially adjacent) 
unevolved clouds are almost unaffected by magnetic fields. 
However, the interaction with the magnetized intercloud gas during the 
pre-collision evolution produces a region of very high magnetic energy  in
front of the cloud. 
In collisions between evolved clouds with transverse field
geometry, this region acts like a 
``bumper'', preventing direct contact between the 
clouds, and eventually reverses their motion. The ``elasticity'', defined as the
ratio of the final to the initial kinetic energy of each cloud, is about 0.5-0.6 
in the cases we considered. This behavior is found  both in adiabatic and radiative cases. 
\end{abstract}


\clearpage

\section{Introduction}

Our understanding of the physical processes of the interstellar medium (ISM) 
of the Galaxy (and of external ones) has progressed tremendously observationally and 
theoretically in the last decade. Part of the required effort has been
stimulated by possible implications for galaxy formation in the early epochs
of the universe, but it is clear that many aspects of the subject pose 
specific physics questions
that are still unsolved and, therefore, interesting to study in their
own right. 

That the ISM of our galaxy should present a multi-phase structure has 
been put forward for more than three decades, in its various versions
including a two-,  three- (and even four-) phase medium. The concept of 
a number of thermal phases coexisting in pressure equilibrium has now been
developed further by Norman \& Ferrara (1996) who found that, once turbulence
is taken into account, a generalization to a continuum of phases is required.

One of the aspects that has been recognized by essentially all the authors
of the above mentioned studies as a crucial physical phenomenon in 
such a multi-phase environment is represented by ``cloud collisions'' (CCs). 
Of course, the term ``cloud'' might literally be appropriate only 
for an approach that is based on a somewhat simplified  thermal 
characterization of the ISM, whereas there is growing evidence that
the dynamics of the gas, either in ordered or random/turbulent form,
could govern its large-scale distribution of the gas. Nevertheless, collisions among
fluid elements, in general, should be quite frequent and common in
the ISM, and without sticking to any particular global model, the first aim
of the present series of papers (Ricotti, Ferrara \& Miniati 1997, RFM; 
Vietri, Ferrara \& Miniati 1997; Miniati \etal 1997, Paper I) is to clarify 
the physics of such events, with special emphasis on the fate of ``clouds''
(\ie, gas clumps) and the dissipation of their kinetic energy. 

The main motivation for this study (and others) is to provide a firm
physical basis for global models of the ISM and galaxy evolution.            
Since the early paper by Oort (1954) several attempts (Field \& Saslaw 1965,
Habe \etal 1981, Struck-Marcell \& Scalo 1984, Ikeuchi 1988, Vazquez \& Scalo
1989, Theis \etal 1992, Jungwiert \& Palous 1996) have been made
to interpret the observed properties of galaxies (as, for example, their 
star formation history, ISM phase evolution and chemical evolution) as
regulated by cloud self-interactions and with the environment (other
phases, radiation field, gravitational potential). 
Basically, the assumption is that the ISM is mostly in cold gas clumps
(\ie clouds) with a spectrum of sizes, which are interacting, coalescing,
forming stars (above a critical mass), fragmented again by energetic 
stellar events and re-injected into the ISM again. The model of Vazquez \&
Scalo (1989), for example, includes heuristic prescriptions on the fate of
collisions through a function $f_c$ of the relative velocity. In this way 
they have been able to explain a number of interesting nonlinear behaviors
as star formation bursts, whose properties are sensibly influenced by
cloud interactions. Similar remarks can be done for models using 
cloud interaction physics to predict the phase interchange in the ISM
(Habe \etal 1981, Ikeuchi 1988). 

Several physical aspects of  cloud collisions have been theoretically investigated in the past
(Stone 1970a,b; Smith 1980; Hausman 1981; Gilden 1984;  Lattanzio \etal 1985; 
Klein, McKee \& Woods 1995); a detailed overview of the characteristics
of \cc is given in Paper I and RFM.  
In spite of the fact that a Galactic collision event is difficult to observe
because of its infrequency, short duration, emission and identification,
a growing amount of  observational evidence for the events is slowly accumulating:
some examples are found in NGC 1333 (Loren 1976), in Heiles cloud 2 
(Little \etal 1978), in Draco (Rohlfs \etal 1989) and in NCP (Meyerdierks 1992).
These detections often correspond to collisions involving clumps 
hosting  a star formation region (Loren 1976). More recently Vall\'ee (1995)
has collected convincing  data for a cloud collision event toward IRAS 2306+1451.

So far only minor attention has been devoted to magnetized \cc. Some 
pioneering analytical MHD work can be found in the literature (Clifford \& Elmegreen 1983) 
but numerical studies have been overwhelmingly limited to hydrodynamical calculations. 
This is surprising, since by now magnetic fields have been detected throughout our galaxy
by a number of dedicated experiments.
Observations (Spitzer 1978, Zeldovich, Ruzmaikin, \& Sokoloff 1983 and 
references therein) suggest that its orientation 
is mainly parallel to the galactic plane and, according to some authors,
it becomes toroidal at high latitudes (Gomez De Castro, Pudritz, \& Bastien 1997).
The magnetic field is further believed to consist of a mean systematic component 
and of a random one. The strength of both is found to be approximately 
a few $\mu$G. The evidence is provided by 
Faraday rotation, synchrotron radiation emitted by energetic
electrons (cosmic rays), starlight polarization and Zeeman effect measurements
(see \eg Zeldovich, Ruzmaikin, \& Sokoloff 1983, for more details).
Direct information about the magnetic field along the line of sight ($B_{\parallel}$)
in galactic diffuse
clouds, is obtained by observing the Zeeman splitting of the 21 cm radio line.
With this technique the magnetic field strength is found to range on average 
between 3 and 12 $\mu$G both in \ion{H}{1} and CO diffuse clouds (Myers \etal 1995). Also
Heiles (1989) finds $B_{\parallel}\sim 6.4~\mu$G observing 
``morphologically distinct \ion{H}{1} shells''. 

Recent measurements carried out by Myers \& Khersonsky (1995) have 
substantially improved our knowledge of the properties of magnetic
fields in interstellar  clouds. 
For the \ion{H}{1} diffuse 
clouds in their sample, $\log x_e$ is found to range between -2.7 and -4.9,
$x_e$ being the electron fraction.
The kinetic Reynolds number, Re (=$vr/\nu$), and even the magnetic Reynolds 
number, Re$_M$ (=$v\ell/\nu_M$) turn out to be large   
for $v\sim$ a few km sec$^{-1}$ and $r\sim\ell\sim 1$ pc. Those results
validate our use of an ideal MHD code (see \S \ref{code}) for these simulations.
Large Reynolds numbers
are very familiar in astrophysics and characterize non-viscous flows. In addition
when Re$_M\gg 1$, the field lines are well coupled with the neutrals and the
magnetic flux is frozen into the fluid. Finally
the ambipolar diffusion time is given by
\begin{equation}
\tau_{AD}=\left(\frac{L}{v}\right) Re_M=2.2\times 10^{10}
\left(\frac{n}{{\rm cm}^{-3}}\right)^2 \left(\frac{L}{{\rm pc}}\right)^2
\left(\frac{B}{\mu{\rm G}}\right)^{-2} x_e~{\rm yr},
\end{equation} 
and therefore the field should not decay through this process, in the 
timescales relevant for the clouds under study.
In particular this ambipolar diffusion should affect the clouds neither during 
their propagation through the ISM prior to, nor during the collisions.

Real structures in the ISM will have complex geometries, so any attempt to
model specific interactions in detail will require 3D simulations. Yet, ours 
is the first explicit MHD study of CC, where the physical effects due to 
the presence of a magnetic field are being investigated.
Because any 3D studies will certainly require interpretation of
very complex patterns and behaviors, we anticipate those works with
an explorative 2D study that should contain many of the same physical behaviors,
and, being far simpler to understand, offers a practical basis for comparison. 
Several important and fundamentally different aspects of the physics of CCs 
are expected to come out of 3D calculations. One important example is the 
appearance of Kelvin-Helmholtz (KHI) and
Rayleigh-Taylor (RTI) instabilities of the cloud surface along the cylinder 
axial direction, which are suppressed in
2D calculations. This might be particularly relevant when the magnetic field is
transverse to the cloud motion, because in this case instabilities on the plane
perpendicular to the cylinder axis can be suppressed by the development of an
intense magnetic field at the cloud nose (Jones \etal 1996). This limits the
time over which the 2D flows are really representative.
At the same time, an initial look 
to our preliminary result of 3D single cloud calculations
(Gregori \etal 1998) reveal that part of the MHD structure relevant to CCs 
developed by three-dimensional clouds is qualitatively similar to that 
seen in 2D clouds. Even though other quantitative differences must occur 
(see \S \ref{sumdis}),
that result certainly supports the validity of our approach consisting of an
initial explorative study of this yet uninvestigated problem. 

Since our objective is the examination of explicit MHD effects in 2D
cloud collisions, we try to follow as closely as possible the 
analogous HD simulations presented in Paper I.
The plan
of the paper is as follows. In Sec 2 we give some general considerations
and introduce the relevant physical quantities of the problem; in Sec. 3
we briefly describe the numerical code and the experimental setup. Sec. 4
is devoted to the results, which are discussed and summarized  in Sec 5.

\section{General Considerations and Parameters of the Problem}

\subsection{Gas Dynamics}
\label{gasd}

In this section we review some basic aspects of purely hydrodynamical CC
(HD CC), a
problem already studied in great detail by previous authors
(\eg Paper I, Klein \etal 1995 and references therein).
The natural timescale for CC is given by 
\begin{equation}
\label{coll}
\coll = \frac{R_c}{v_c},
\end{equation}
which is approximately the time required for the shock generated by the collision 
to propagate across the cloud radius, $R_c$. We suppose that 
clouds initially have a circular cross section.
The main parameter determining the character of non self-gravitating HD CCs is
(Klein \etal 1995)
\begin{equation}
\label{eta}
\eta=\frac{N_{rad}}{n_c R_c},
\end{equation}
where $N_{rad}=n_cv_c\rad$ is the radiative cooling column density through 
one cloud, $n_c$ is the cloud number density, and
$\rad $ is the cooling time (Spitzer 1978). Combining Eqs. \ref{coll} and 
\ref{eta} we have $\eta=\rad/\coll$. Accordingly,
 if $\eta\le 1$ significant radiative cooling takes place during the collision;
when $\eta\gg 1 $ emission processes become
unimportant and the flow behaves adiabatically. In Paper I we concluded, after 
several low resolution tests, that the latter condition can be well represented
by the weaker relation  $\eta > 1. $ Once the two phase model for the ISM 
is assumed, for a given cloud velocity these conditions imply that 
impacts between
larger clouds are more influenced by radiative cooling (RFM, Paper I).

Since much of the basic physics of \cc has been explored using head-on events,
which can provide a standard for comparison, in this paper we focus entirely 
on head-on \cc.
In general, head-on symmetric HD \cc evolve through four main phases 
(Stone 1970a, 1970b); namely, {\it compression, re-expansion, collapse} 
and under some circumstances {\it dispersal}
(Paper I). The occurrence of these four phases was first pointed out by Stone 
in his pioneering work (1970a, 1970b). It was also subsequently confirmed
by high resolution hydrodynamic calculations (Klein, McKee \& Woods 1995,
Paper I), which allows for the highest resolution. Similar results were also 
found 
through SPH calculations (\eg Lattanzio \etal 1985, Lattanzio \& Henriksen
1988). However, the limitations of these calculations, both in resolution and
in the SPH method of simulation, have led to some misinterpretations of the
physics of CC, like the hypothesis of ``isothermality'', 
as discussed in Paper I.
To a certain extent the same depiction of four phases can be drawn for 
magnetohydrodynamical \cc (MHD CCs) as well, although as we shall see, 
there are some important differences. 
However, we adopt this general terminology as a useful tool to refer to the 
various stages of the \cc.

\subsection{Cloud Propagation through a Magnetized ISM}
\label{desp}

The HD of a dense cloud moving into a low
density medium has been central to the study of several authors 
(Jones \etal 1994, 1996; Schiano, Christiansen \& Knerr 1995; Murray \etal
1995; Vietri \etal 1997; Malagoli, Bodo \& Rosner 1996) 
who generally concentrated on the growth of KHI and
RTI in such conditions. We refer the interested
reader to these works for an exhaustive description of this topic.
When considering the motion of a cloud through a magnetized medium,
new parameters, in addition to those introduced so far (\S \ref{gasd}),
must be considered.
In particular, the initial magnetic field is completely defined by its strength and
orientation. The former parameter is usually expressed in terms of
\begin{equation}
\label{beta}
\beta_o = \frac{p_g}{p_B} = \frac{2}{\gamma}\left(\frac{M_A}{M}\right)^2,
\end{equation}
where $p_g$ and $p_B=B^2/8\pi$ are the gas and magnetic pressure, 
and $M=v_c/c_s$ and $M_A=v_c/(B/\sqrt{4\pi\rho})$ are the sonic and 
Alfv\'enic Mach number respectively.
The magnetic field orientation is, in general, determined by two angles;
in 2D simulations, with the field lying in the computational plane,
they reduce to $\theta$, the angle between the
cloud velocity and the field lines. 
As long as the unperturbed magnetic field is dynamically unimportant
($\beta_0 \gg 1$) 
the initial evolution of a MHD cloud is similar to a HD one.
This is the case we consider below, where we adopt $\beta =4$.
So, in the presence of a weak field, a stationary bow shock develops
on a timescale $\tau_{bs}\sim 2\coll$. Further, a ``crushing'' shock is 
generated and propagates through the cloud with relative speed $v_{cs}\simeq 
v_{c}/\chi^{1/2}$, on a timescale ({\it crushing time})
\begin{equation}
\tcr = \frac{2R_c}{v_{cs}}=\frac{2R_c\chi^{\frac{1}{2}}}{v_c},
\end{equation}
where $\chi=\rho_c/\rho_i$ is the ratio of the cloud and intercloud 
medium densities.
Finally a low pressure region (wake) forms at the rear of the cloud and,
interacting with the converging flow reflected off the symmetry axis (X-axis), 
generates a relatively strong tail shock. 
However, over time, new features develop in response to the magnetic field. 
Jones \etal (1996), in a 2D study of individual MHD supersonic clouds,
identified several of these for an  adiabatic, 
high Mach number cloud with modest density contrast, $\chi=10$.
Mac Low \etal (1994) also studied the MHD evolution of 2D individual, shocked clouds, which 
behave in a qualitatively similar manner.
Utilizing these works, in the next paragraphs we give a brief review  of the 
main features that result from inclusion of a magnetic field during the interaction with the 
intercloud medium. 

First, the orientation of the magnetic field with respect to the 
direction of the cloud motion is particularly important. So far only
two extreme cases have been published; namely, cloud motion parallel 
({\it aligned}) or perpendicular ({\it transverse}) to the initially uniform field.  
We present elsewhere calculations with oblique magnetic fields 
(Miniati, Jones \& Ryu, 1998).
In the aligned case, the field lines, following the flow, are swept over 
the cloud. As a result, those lines anchored at the cloud nose, are 
pulled, stretched and folded around the cloud. 
Eventually, these lines experience magnetic reconnection, forming new flux 
tubes passing around the cloud, somewhat like streamlines in a smooth flow.
In this region the magnetic field never becomes dynamically dominant, although
its realignment around the cloud contributes to smoothing and, therefore, 
stabilizing the flow. On the other hand, field lines are drawn into the cloud 
wake. Flow in the wake stretches lines
anchored in the cloud material, causing the intensity of the magnetic field
to increase. This feature, referred to as the post-cloud ``flux rope'' 
(Mac Low \etal 1994, Jones \etal 1996), is the only one that becomes 
magnetically dominated ($\beta\gtrsim 0.1$) for an aligned geometry.
A similar wake region forms also in the transverse field case, where 
the field lines drape over the cloud, converging in its wake. In this 
case, however, those
field lines are anti-parallel across the symmetry axis.
Above and below the symmetry axis the field structures in the wake
initially are relatively uniform with a very sharp transition  between
them corresponding to a thin current sheet. Classically such a thin 
current sheet is unstable to the
resistive, ``tearing-mode'' instability (\eg Biskamp 1993, p. 73), in 
which the current sheet breaks into line currents and the magnetic
field reconnects across the sheet. 
The instability condition is that the thickness of the sheet is
much smaller than its width (\eg Biskamp 1993, p. 152 for details).
Indeed, we see these sheet transitions break up into a series of closed field
loops that are the signature of this instability (\eg Melrose 1986, p 151).
This rapid modification of the magnetic field topology is often called 
``tearing-mode reconnection'' (\eg Melrose 1986) and typifies the 
reconnection that occurs in our simulations.
A very clear illustration of the evolution of one such example is
shown in Miniati \etal (1998, Figure 3).
Because of this behavior the magnetic field intensities in the transverse
case wakes are lower than in the aligned case.
On the other hand, the field lines in front of the cloud 
are compressed and, more importantly, 
stretched around the cloud nose. In this region, unlike the previous aligned
field case, 
reconnection does not occur for these field lines. Therefore,
the magnetic field becomes very intense ($10^{-2}\le \beta \le 10^{-1}$) on the
cloud nose,
forming a {\it magnetic shield}. As pointed out by Jones \etal (1996), 
the main reason of magnetic energy enhancement is the stretching of 
the lines as these are swept up by the cloud. 
The timescale for the growth of the magnetic energy is given in eq. 9 of 
Jones \etal (1996), in the spirit of a first order approximation quantity, as
$\tau \sim v_c R_c^{-1} = \coll$. Soon however, nonlinear effects become important
and a more realistic timescale, as long as 2D approximation is valid,
is provided by (Miniati, Jones \& Ryu 1998)
\begin{equation}
\label{rate}
\tau \sim (\beta\chi)^{2/3} M^{4/3}~\coll.
\end{equation}
Further details on the development of the magnetic shield and on its dependence
on cloud characteristics can be found in Miniati, Jones \& Ryu (1998).
It is important to notice that the magnetic shield 
acts to prevent the growth of KHIs and RTIs on the cloud surface.

When radiative losses are included, as in the HD calculations, 
the thermal energy of the compressed gas is lost, reducing its pressure and 
allowing the cloud material to be compressed to very large densities. 
Particularly in the transverse field case, the cloud aspect ratio (length, $x$,
to height, $y$) is highly increased by this effect. 

\section{Numerical Setup}
\subsection{The Code}
\label{code}

Our \cc simulations have been performed using an ideal MHD code,
based on a second order accurate, conservative, explicit TVD method.
Details of the MHD code 
are described in Ryu \& Jones (1995) and Ryu, Jones \& Frank (1995);
a brief description of some aspects concerning the inclusion of 
cooling and of a mass tracer is given in Appendix A.
The $\nabla\cdot{\bf B} = 0$ condition
is maintained during the simulations by a scheme similar
to the Constrained Transport (CT) scheme (Evans \& Hawley 1988; Dai \&
Woodward 1997),
which is reported in Ryu \etal (1998).  We have used the 2D,
Cartesian version of the code.  The computational domain is on the $xy$
plane, and the Z-components of velocity and magnetic field have been
set to zero.

\subsection{Grid, Boundary Conditions and Tests}

In each CC simulation only the plane $y\ge 0$ is included
in the computational box, and reflection symmetry is assumed across the X-axis.  
The length scale is chosen for each case so that $R_c = 1.0$ and
the computational domain is adjusted to minimize boundary
influences, as listed in Table \ref{tabcs}.
Since the grid is Cartesian, our clouds are actually cylinders,
with axes in the Z-direction. 
The top and right boundaries are always open.
The left boundary is open in the asymmetric cases (3 and 4 of Table \ref{tabcs})
and is reflective (as the bottom boundary) otherwise, when there is a
mirror symmetry to the collision.
With this latter choice we are allowed to use only half of the grid and reduce 
the computational time of the calculation. 
Only the highest resolution calculations, characterized by 50 zones across 
the initial cloud radius, are presented here. Lower resolution 
(25 zones per initial cloud radius) tests were also performed in order to check
for consistent behavior. It turns out that beside inevitable quantitative 
differences, for each case we studied there is absolute 
consistency in the CC outcome. In particular it is apparent the qualitative 
agreement  between low and high resolution calculations, 
in the density distribution and magnetic field structures that form 
out of the CC. 

\subsection{Initial Conditions}
\label{param}

In this section we will discuss the initial conditions for our CCs.
It is worth pointing out from the very beginning that, following the results of
Paper I, most of the simulated CCs involved {\it evolved} clouds, \ie clouds
that have propagated through the ISM for about $\tcr$ before colliding. 
Initially, individual clouds have a circular cross section and uniform 
density and are in pressure equilibrium with a uniform background medium;
the magnetic field is also assumed uniform throughout the domain. 
The relevant parameters, whose numerical values are given below,
are the density contrast $\chi$, the Mach number $M$ and the cloud radius $R_c$.
The exact thermodynamic quantities characterizing the initial equilibrium 
state are not particularly important as their memory is lost soon after the 
beginning of the cloud evolution:
for a supersonic motion, the thermal pressure of the shocked 
gas and the ram pressure of the flow are dynamically far more important than 
the initial pressure balance. In addition, as
already pointed out in \S \ref{desp}, the cloud motion through the intercloud 
medium produces a variety of features which strongly alter the initial 
configuration. 
As a result, despite the simplicity of the conditions at the onset of the
cloud motion, before the collision takes place 
both the gas and the magnetic field are characterized by a rich structure. 
A comparison with the collision of two unevolved clouds is provided 
in the results section (\S \ref{by0}), to show the importance of considering 
prior cloud evolution in this study. 

The initial values of the above parameters are the same as in Paper I
and are listed below. These values are inspired by 
the most recent observational studies
and thought to be representative of the magnetized ISM.
We assume a specific heat ratio $\gamma=5/3$   
throughout. The inter-cloud medium has a density $n_i=0.22$ cm$^{-3}$
and temperature $T_i=7400$ K.
Clouds are characterized by a
density contrast $\chi=n_c/n_i$=100, so that the cloud density and temperature
are $n_c = 22$ cm$^{-3}$ and $T_c = 74$ K respectively and 
$\rad \approx 3.7\times 10^4$ yr. 
The sound speed in the inter-cloud medium turns out $c_{si}\approx 10$ 
km s$^{-1}$ and the equilibrium thermal pressure for the ISM
$p_{eq}/k_{\tiny B}=1628$ K cm$^{-3}$. 
When initially set in motion parallel to the X-axis, each
cloud has a Mach number $M=v_c/c_{si}=1.5$, and therefore
$v_c\approx$ 15 km s$^{-1}$. 
Setting $R_c = 0.4$ pc, yields
$\tau _{coll}=R_c/v_c \approx 2.6\times 10^4$ yr. Since this implyes 
$\eta=\rad/\coll \approx 1.4 > 1$, 
in accord with \S \ref{gasd} these collisions behave 
adiabatically and therefore the cooling can be turned off. In the 
following collisions involving these clouds are referred to as adiabatic
cases. On the other hand, setting $R_c = 1.5$ pc
we have: $\tau_{coll} \approx 9.7\times 10^4$ yr
and therefore $\eta\approx 0.38$. These are the radiative cases.
Table \ref{tabca} summarizes these cloud characteristics.

With a magnetic field included, three new parameters with respect to the HD case have
to be determined: the field orientation and strength.
The orientation is determined in 2D by a single parameter, namely  
the angle between the initial cloud velocity and magnetic field. 
We explore two cases: magnetic field parallel (aligned case) and 
perpendicular (transverse case) to the cloud velocity. 
For the initial strength of the magnetic field,
conveniently expressed by the parameter $\beta_0$ (see Eq. \ref{beta}),
we assume $\beta_0 = 4$, corresponding to $B=1.2~\mu$G. 
It could be argued that this value is somewhat smaller
than what is usually observed. However, during the following cloud evolution, 
the field is stretched and amplified and in several regions
becomes energetically dominant ($\beta \ll 1$ and $B>1\mu$G, \S \ref{desp}). 
In the resulting configuration, therefore, the magnetic field influence
is not highly sensitive to this choice.

Initially, the Jeans length of our typical diffuse cloud is 
$\lambda_j\approx$ 29 pc$\gg R_c$. Even though large density enhancements are 
produced during the compression phase in symmetric radiative collisions,
$\lambda_j$ never becomes smaller than the vertical size of the clouds.
For this reason we have neglected self-gravity throughout our calculations 
(see also Klein, McKee, \& Woods 1995). Since we have concentrated on diffuse
clouds, as opposed to molecular complexes, this approximation is
justified.

Similarly to Paper I, we consider  collisions between 
both radiative and adiabatic clouds. However, for brevity,
we discuss here only the most significant new results
for the two different 
cases. Table \ref{tabcs} summarizes the parameters of  the
collisions discussed below.
For Cases 1 and 2 in Table \ref{tabcs},
the cloud begins its independent evolution at
 $t = -\slantfrac{3}{4}\tau_{cr}$,
where $t=0$ corresponds to the instant when the bow
 shock of the two clouds first touch.
Analogously, for Cases 3 and 4 the two clouds are placed on
 the grid at $t = 0$ with 
their bow shock next to each other, after evolving for
 $\slantfrac{1}{2}\tau_{cr}$
and $\tau_{cr}$ respectively.
In these four cases the initial magnetic
field was aligned with the cloud motion, whereas for Cases 5-7 it
was transverse.
Case 5 is the only non-evolved calculation we present:
by that we mean a uniform cloud of circular cross section
placed on the grid in such a way that its
boundary is only 2 zones from the (reflecting)  Y-axis at $t = 0.0$.
Finally in Cases 6 and 7 the cloud starts its evolution at
$t = -\slantfrac{3}{4}\tau_{cr}$ and then is treated as in Cases 1 and 2 
respectively. 
Animations of each simulation have been posted on our
World Wide Web site at the University of Minnesota.

\section{Results}
\label{resu}
\subsection{Aligned Field (BX, Cases 1-4)} \label{bx}

In the aligned field case 
MHD CC show many similarities with HD CC. These are illustrated in
Figure \ref{plot-bx}, where the evolution of various MHD cloud integral 
properties are plotted as a function of time. The curves of kinetic and 
thermal energy as well as \ycm closely resemble those for HD clouds in
Figures 8 and 9 of Paper I and the occurrence of all the phases 
characteristic of a HD CC (see \S \ref{gasd} and Paper I) is clearly seen.
In general a stronger compression is generated when 
the clouds have mirror symmetry across the impact plane and 
only a weak reexpansion takes place in radiative cases, 
because most of the thermal energy is radiated away. 
The top right panel of 
Figure \ref{plot-bx} displays the total magnetic energy.
Since its variations are related to compression and/or stretching of the field 
lines, this quantity gives an approximate measure of the overall interaction between 
the magnetic field and the gas. In the adiabatic cases (solid and dotted lines) 
the large expansion undergone by the 
cloud gas produces both significant stretching and
 compression of the field lines.
As a result, during the re-expansion phase 
the total magnetic energy increases of about 30\%. 
On the other hand, in all radiative cases the total magnetic
energy suffers only slight variations.

We now begin specific comparison, considering the adiabatic collision of two 
evolved identical clouds with aligned fields. 
 
\subsubsection{Symmetric Cases}
\label{bxsy}

The evolution of Case 1 is reported in Figs. \ref{bxadsy}a and \ref{bxadsy}b, 
representing the field line geometry (left panels)
and density distribution (right panels) at four different times 
($t/\coll=$4.5, 24, 48.75 and 75).
As complementary
quantitative information, Figs. \ref{rho-cut}, \ref{pre-cut} and 
\ref{bpr-cut} provide, for the same times, cuts along the primary axes of
density, thermal and magnetic pressure, respectively. Lines there 
correspond to $t=4.5\coll$ (solid), $t=24\coll$ (dotted), 
$t=48.75\coll$ (dashed) and $t=75\coll$ (dot-dashed).
Cuts in the top panels are along
the X-axis whereas those in the bottom ones are along the Y-axis.

As already suggested above, the evolution
of the CC is substantially unaffected by the magnetic 
field. Indeed the CC goes through 
the four phases mentioned in \S \ref{gasd} as in the HD case. For the 
compression and reexpansion phase there is a close quantitative 
correspondence, as it can be inferred from a comparison of
Figs. \ref{rho-cut} (solid and dot line) and Figure 2 of Paper I (although 
lines in the two figures do not correspond exactly to the same time). 
This is not unexpected, though, 
since the compression and reexpansion phases are dominated by the 
high pressure of the shocked gas. 

During the reexpansion phase a thin shell of dense ($4\roi\le\rho\le10\roi$)
cloud gas forms behind the reverse shock of the expanding material. A long 
finger appears on the X-axis, due to the fact that the reexpansion finds an 
easy way through the flux rope, where the density and the pressure are 
quite low (Mac Low \etal 1994, Jones \etal 1996). Also, in contrast to the HD 
case, the shell boundaries in Figure \ref{bxadsy}a (bottom left)
are quite sharp. This difference has an important 
physical base. In fact, although the initial magnetic field has not sufficient 
strength to inhibit the onset of KHI (Chandrasekhar 1961), nevertheless
it is able to reduce and eventually stop its growth. In fact, 
as the magnetic field lines, frozen in the gas, get stretched in the turbulent 
flow, their strength is increased until during eventual reconnection they 
redesign the flow
pattern to a more stable configuration. The criterion for this field dominance
is that the local alfvenic Mach number falls to order unity or less
(Chandrasekhar 1961, Frank \etal 1996, Jones \etal 1997). 
This is evinced by the presence of several field line loops on the 
external side of the gas shell
(lower left panel in Figure \ref{bxadsy}a). 
Inside the shell the magnetic field intensity has 
severely dropped. According to Figure \ref{bpr-cut} 
the magnetic energy density (dotted line) has been reduced with respect to 
its initial value (solid line) by a factor ranging from $10^2$ to
$10^4$. This cannot be accounted for by expansion alone; indeed complex
reconnection processes at the beginning of the reexpansion phase
are responsible as well. 

The collapse phase (at $t\sim 18.75\coll$) is chaotic and turbulent: in and
around the cloud gas
the density distribution is rather clumpy and the magnetic field 
has a tangled structure (top panels of Figure \ref{bxadsy}b and dash lines
in Figs. \ref{rho-cut} and \ref{bpr-cut}). This situation will persist until 
the end of the simulation (dot-dash lines
in Figs. \ref{rho-cut} and \ref{bpr-cut}). It is during this phase that the 
magnetic field produces a qualitative change in the evolution of the CC. 
In Paper I we showed that during the HD collapse 
phase the reverse shock propagates toward the inner 
region of the expanded cloud gas, whereas the external layer
is shredded by KHIs and RTIs. At the end numerous filaments 
fill a region of about the same size and shape
of the shell at its maximum extent.
In the present case, on the other hand, the magnetic field 
lines which have been stretched by the vertical expansion of the gas shell, 
begin to relax, accelerating the gas at the top of the shell (near the Y-axis)
toward the X-axis (top panels of Figure \ref{bxadsy}b). 
At the end of the collision ($t=75\coll$)
the initial cloud material is confined in a layer beneath the relaxed 
magnetic field lines, with a mean density
$\rho\sim 4\roi$ (bottom panels of Figure \ref{bxadsy}b) and is 
still laterally expanding.  
Within this new structure, with a thickness of several initial cloud radii, 
we find a clumpy 
density distribution (dot-dash lines in Figure \ref{rho-cut}) and a weak,
irregular magnetic field (dot-dash lines in Figure \ref{bpr-cut}). Outside it,
on the other hand, the magnetic field has the same initial configuration but
greater strength, often twice as much as its initial value
(dot-dash line in bottom panel of Figure \ref{bpr-cut}).

The analogous radiative case (Case 2) is illustrated in Figure \ref{bxrasy}
and its quantitative properties plotted in  Figure \ref{cutx-rad}.
Again, an aligned magnetic field does not seem to influence the 
collision.
As in the HD case efficient 
radiative cooling allows a much higher gas compression (solid line in 
top panel of Figure \ref{cutx-rad}) and inhibits the strong reexpansion that 
would be driven by the high pressure of the shocked gas (solid line in 
bottom panel of Figure \ref{cutx-rad}). During these phases the clouds coalesce,
generating a well defined high density round structure 
about twice as large as the initial single cloud 
(this can be inferred from the position of 
the sudden drop in the dot line in 
top panel of Figure \ref{cutx-rad}). No collapse phase ever happens. 
In addition a narrow jet of gas, characteristic also of radiative HD CC, 
is formed and extends along the Y-axis. As it propagates transverse to 
magnetic field lines a high density spot is created at the leading edge.
However those features are usually unimportant because they only involve a
negligible fraction of the total mass. Of more interest is instead the final 
fate
of the mentioned structure. At the end of our simulation ($t=30\coll$) its 
edge near the X-axis is still expanding along the X-axis with a speed 
$v\sim 0.1 c_{si}$. According to the three density cuts in the top panel of 
Figure \ref{cutx-rad} the cloud edge (located at the sharp drop in each line) 
has been 
expanding at roughly the same speed ($\sim 0.1 c_{si}$) throughout the 
evolution. Based on this velocity we can, therefore, estimate the time 
$\tau_\alpha$ for the
density of the new structure to drop by a factor $\alpha$. Assuming that only 
one dimension of the cloud volume increases (at the speed of 0.1 $c_{si}$)
as a result of its expansion, we have $\tau_\alpha\simeq \alpha R_c/(0.1 c_{si})
= \alpha 10 M \coll =  \alpha ~1.5 \times 10^6$ yr
(we used $\coll\sim 10^5$ yr from \S \ref{param}).
Since inside the new structure $\rho\sim 40-80$, $\alpha$ can be 
as large as 5-10, and $\tau_\alpha$ comparable with the time between two 
cloud collisions. 

\subsubsection{Asymmetric Cases}
\label{bxas}

As in Paper I we produce a simple asymmetric CC by colliding clouds with the same 
initial mass, velocity and radius, but evolved individually for a different 
time interval before they collide. By contrasting with mirror symmetric
cases we can begin to see properties that are symmetry dependent. 
In the following cases (Case 3 and 4)
the two clouds have been evolved for about $\slantfrac{1}{2}\tcr$ and 
for $\tcr$ respectively (Table 1).

The evolution of the adiabatic asymmetric CC resembles the analogous symmetric
Case 1. Figure \ref{bxadas} shows the field lines (top panel) and the density 
distribution (bottom panel) at the end of the re-expansion phase 
($t=22.5\coll$). We can identify a relatively dense shell ($\rho \sim 5\roi$) 
with a large clump of gas on top of it as well as
loops of magnetic field lines generated by reconnection events.
Some new features appear, however, 
as a result of the broken symmetry. An example is the long tail of cloud gas
on the right hand side of Figure \ref{bxadas}, due to the unbalanced momentum 
distributions in the two clouds along the X-axis. Nevertheless they are not
so significant as to alter the general character of this CC with respect to Case 
1. Therefore we expect that the collapse phase and the remaining following 
evolution of this case will not differ qualitatively from that of Case 1.
We also point out that as long as the asymmetry does not prevent the development
of a reexpansion phase, Case 1 can be considered as qualitatively well 
representative of adiabatic MHD CC with magnetic field aligned to the initial
cloud motion.

The radiative case (Case 4), shown in 
Figs. \ref{bxraas}a, \ref{bxraas}b, presents new and
interesting insights.  The most crucial part of the evolution
is represented by the cloud interaction during the compression phase. 
During that phase the older and more compact cloud (C2), moving from the right,
attempts to plow through the other cloud (C1) (Figure \ref{bxraas}a --
$t=6.75\coll$). Some of the features that are visible in Case 2, such as the 
vertical jet of gas, can also be recognize here if one carefully 
accounts for the distortions due to the asymmetry. 

However, we give particular attention to the new 
feature on the X-axis toward the left of Figure \ref{bxraas}a (X$\simeq -5R_c$). 
This is 
a compact clump with density about $5\times 10^2\roi$ and 
velocity along the X-axis, $v_x\sim -\slantfrac{1}{2} c_{si}$.
At the end of the simulation ($t=22.5\coll$) it
has expanded and is still moving toward the left along the X-axis
(Figure \ref{bxraas}b). At this time the 
``mass tracer'' variables allow us to conclude  
that, despite the large prevalence of gas from the more compact cloud (C2),
the new clump is the result of a partial 
coalescence of the two initial clouds.
Its mass is about 10\% larger than the initial mass of
either cloud. Its density varies
between 15$\roi$ and 100$\roi$ and the velocity pattern suggests that it 
is expanding along the X-axis with $v\sim c_{si}/6$. By the same 
argument at the end of \S \ref{bxsy} we can, therefore, conclude that 
for this case $\tau_\alpha\simeq \alpha ~9 \times 10^5$ yr.
Again, before the cloud disperses in the background medium,
the newly-formed clump is likely to undergo another CC. 
The remainder of C1's gas (feature to the right), moving transversely to the 
magnetic field, has formed a long filamentary structure
and has created a sharp cusp in the field lines (Figure \ref{bxraas}b).

This result strongly differs from the analogous 
purely HD calculation, where we found that the collision produced a large, 
low-density-contrast filamentary structure eventually fading into the 
background gas.

\subsection{Transverse Field (BY, Cases 5-8)}
We start this section by presenting the case of a CC between two unevolved
clouds (Case 5). This calculation is mostly intended to provide a reference case
when studying evolved \cc  and, thus, to emphasize the importance of initial 
conditions in calculations of this type. 

\label{by}
\subsubsection{Unevolved Adiabatic Collision}
\label{by0}
The results of this calculation (Case 5) are shown in 
Figure \ref{byad0}, where two density images are superposed 
on field lines. The left panel captures the reexpansion phase at
$t=7.5\coll$; it closely resembles the analogous HD Case 1 of Paper I and
no significant difference from Case 1 (of this paper), where the magnetic field 
was aligned to the motion, can be pointed out. This all means that no major 
role is played by the magnetic field, as it has been the case so far for all
adiabatic CCs. The shell density has typical values ($\sim 10 \roi$) and the
thermal pressure drives the reexpansion. RTIs develop on the shell surface
near the Y-axis. However, 
at the end of the collision a
thick layer, qualitatively similar to that 
formed in Case 1 and 3 forms along the Y-axis. 
We also point out that the evolution of the total magnetic energy (dash line
in bottom left panel) is reliable only up to 
$t\sim 9\coll$, which marks the exit of the forward 
blast shock from the right boundary of the grid, along with consequent outflow
of magnetic energy. 
The right panel image is from $t = 30\coll$. RTIs, aided by the
formation of a slow, switch-off shock, which aligns the field with
the expansion of the shell, have formed a large finger expanding
almost parallel to the Y-axis.

\subsubsection{Evolved Symmetric Cases}
\label{bysy}
As already mentioned in \S\ref{desp} and discussed by 
Jones \etal (1996), an individual cloud
moving transverse to a magnetized intercloud medium develops a region 
of strong magnetic field known as the magnetic 
shield. That
feature dominates the collisions of such evolved clouds from the start
of their encounter.
Consequently, magnetic field effects dominate the evolution in Cases 6 
and 7, in striking contrast to Case 5.
Consequently, it is clear that meaningful simulations of CCs may
depend on understanding the field geometry in their surroundings and
on allowing self-consistent magnetic structures to evolve before
collisions take place. 

Case 6 is shown  in Figs. \ref{byadsy}a and \ref{byadsy}b, both including
field line geometry and density images superposed on the velocity field.
In addition, the solid lines in Figure
\ref{plot-by} show the usual time evolution of integral quantities.
Peaks in the thermal energy and corresponding valleys in the $Y_{cm}$ curve,
are signatures of the bow shock precompression and the ``collision'' 
undergone by the cloud .
Note the simultaneous kinetic energy decrease and
the magnetic energy increase, primarily due to cloud 
interaction with the magnetic field (Miniati \etal 1998) at the 
beginning of the simulation and to the ``collision'' event later on.

As the cloud approaches 
the collision plane (Y-axis) in Figure \ref{byadsy}a (bottom panel), 
the field lines are highly compressed (top panel)
generating a strong repulsive force, opposite to the cloud motion.
Eventually all the cloud kinetic energy is converted into magnetic form
and stored as magnetic pressure. Then the cloud stops and, as the field lines
reexpand, is accelerated backward and its motion reversed
(Figure \ref{byadsy}b). Therefore we can state that
the magnetic shield acts almost like an elastic magnetic bumper.
At the ``apex'' of this reversal phase both thermal
and magnetic energy peak, whereas the kinetic energy obviously is about null
(Figure \ref{plot-by}). The latter, however, at the end of the simulation 
returns to about 60\% its initial value (Figure \ref{plot-by}) and the cloud 
velocity is $v_x\sim c_{si}$. Also thermal and magnetic energy are back to 
their initial values. Therefore the only effect of the collision is to 
dissipate part of the cloud kinetic energy with no other major consequences.

The same qualitative result is also obtained in the radiative Case 7,  
shown in Figs. \ref{byrasy}a and \ref{byrasy}b and in Figure 
\ref{plot-by} (dot line). Whereas the kinetic and magnetic 
energy evolve as in Case 6, the thermal energy and \ycm 
parameter have now similar qualitative behavior but much lower values 
(the scale for the dotted lines in the two bottom
panels of Figure \ref{plot-by}, is 10 times smaller than for the solid ones); 
thus, the evolved cloud is mostly supported by magnetic pressure. 
Figs. \ref{byrasy}a and \ref{byrasy}b show the magnetic bumper effect for 
the radiative case as well. The final kinetic energy is about the same
as in the previous case (50\% of the initial value).
However, the cloud gas has extensively spread out, 
especially in the X-direction, creating a dense elongated structure. 
Because of the high density reached by the cloud during the reversal phase, 
the resulting cloud shape is probably limited by the diffusivity of the code. 
Nevertheless, the final outcome is different from the analogous HD and previous
aligned field cases and is more similar to Case 6. 
The cloud neither disperses nor coalesces; its structure is, however, 
strongly distorted and its kinetic energy partly conserved. 

\section{Summary \& Discussion} \label{sumdis}

We have investigated the role of the magnetic field 
in CCs, through high resolution, fully MHD 2D numerical simulations.
This paper is an extension of the gasdynamical study 
presented in Paper I. Our aim is to provide a first step toward the
understanding of the physical role of magnetic fields in interstellar diffuse \cc. 
In particular, we have studied magnetic influences on: (i) the final fate of the
cloud after the collision (\ie dispersal, coagulation, shattering, filamentation);
(ii) the evolution of cloud kinetic energy and (iii) the 
effects of \cc on the magnetic field structure in and around the clouds.
These simulations represent only an initial attempt to study a 
complex problem. To identify the most obvious and simplest behaviors
we have restricted the geometrical freedom of the flow to 2D.
For all behaviors, but particularly for (iii),  our results need to be 
confirmed  by more thorough and extended 3D calculations.
The main results can be summarized as follows:

\begin{itemize}
\item Adiabatic, aligned field \cc are disruptive (as in the HD case), both
for symmetric and asymmetric events. The remnant consists of an elongated 
structure of low magnetic energy in which cloud and background gas are mixed
together.

\item Addition of an aligned field to radiative, symmetric \cc does not
change the fact that they dissipate most of the cloud
kinetic energy, which leads to almost complete coalescence of the two clouds. 
During asymmetric collisions of this type
coagulation takes place as well, but the final structure has a mass 
only slightly ($\sim$ 10\% for Case 4) 
larger than either cloud initial 
mass; little alteration of the magnetic field line pattern is seen. 
This result is important, since purely HD asymmetric collisions of diffuse clouds have 
been shown to be highly disruptive (Paper I, 
Klein \etal 1995 and references therein).

\item In 2D motion of clouds moving transverse to the magnetic
field leads to the formation of a magnetic shield in front of each cloud. 
When two evolved clouds of that kind run into each other, a magnetic
shield may prevent direct collision
from taking place. In our simulations the clouds remain
separated by a magnetic barrier and bounce back with 
a fraction $\epsilon$ of the initial
kinetic energy. According to our results $\epsilon\sim 0.5-0.6$ for both
the adiabatic and radiative cases. This is probably an upper limit in
more realistic situations, including 3D, and especially off-axis
collisions, since the magnetic bumper may be less developed and other
degrees of freedom (\eg rotation) are available.In addition, if a third
dimension were included, after the collision of the clouds the compressed
magnetic shield would partially reexpand perpendicular to the direction of
the initial motion, thus reducing $\epsilon$ value further.

\end{itemize}

Despite the caveats mentioned, much of the character 
represented in the last bullet may be independent of the 
symmetry of the collision. This was tested in part through a low resolution
2D numerical experiment of an asymmetric head-on, transverse-field collision. 
Results were consistent with those cited. In general,
the magnetic shield is expected to work at some level for off-axis
and largely asymmetric cases; transfer of momentum 
and angle scattering would occur in a similar way as in head-on collisions. 
When the magnetic field is aligned to the cloud motion,
some of the arguments discussed in Paper I for off-axis HD CC should apply 
here as well.
In particular off-axis CCs with an impact parameter $b\ll R_c$
should be well represented by our asymmetric cases. 
Also for adiabatic cases,
we expect that even for $b\gtrsim R_c$, CCs should produce a reexpansion
that is strong enough to disperse the clouds. The asymmetric 
radiative case results, however, caution from extending the HD results to 
radiative MHD CC when $b$ is comparable to $R_c$. Those cases must be investigated
in the future. Finally for $b\sim 2R_c$ the collision should produce only
minor perturbation to the clouds. 

A striking difference exists between the outcome of \cc when aligned and 
transverse field geometry are considered. Therefore, in order to model 
correctly \cc in the ISM it is important to understand 
the conditions  for the formation of the magnetic bumper.
Two points are crucially important in this regard:
(i) the assumed cloud shape and (ii) the initial configuration. 

As for the former, no 3D MHD individual supersonic cloud numerical 
simulation has been published 
so far. Especially near the nose of the clouds we should expect to see some
kind if magnetic shield develop. On the other hand divergence of the
flow away from the nose and transverse to the prevailing field
orientation should advect field lines away from the nose, thus
limiting its development and extent. The importance of that effect
will depend on the geometry of the cloud. A ``pointed cloud'' will
have only a very limited shield.
For example, Koide \etal (1996), studying the propagation of extragalactic 
jets through a medium with an oblique magnetic field, find that the field lines 
distort to let the jet pass through. 
This effect reduces the strength of the magnetic shield and could be important
for spherical clouds. However, for cylindrical or filamentary clouds, with axes
transverse to the motion, field lines may be trapped at the front of
the cloud long enough to play a major dynamical role. 
We point out that elongated clouds are not simply
convenient to our 2D approximation.
Rather, elongated shapes are expected for clouds that form in 
a magnetized environment, where the support provided by the magnetic 
pressure is anisotropic (\eg Spitzer 1978). 
As already pointed out by Jones \etal (1996), realistic clouds show 
strong shape irregularities where the field lines can penetrate, 
be captured and, therefore, stretched to form some sort of magnetic bumper. 
But even if not so, when a cylindrical cloud moves in a transverse magnetic 
field, the motion of the gas along the cloud major axis, away from the 
stagnation point, is certainly slower at the center of the cloud where the
stagnation point is located, than near the sides. Since the field lines 
are frozen into the gas, the central region of 
the cylindrical cloud is where the field 
lines are held longest. Therefore, an uneven magnetic tension is applied on
the cloud and a bending of the latter is produced in the central region.
As a result it becomes more difficult for the field lines to slip by 
the cloud enhancing the deformation of 
the cloud and increasing trapping of the field lines.

Finally we have just begun a set of preliminary, low resolution numerical 
calculations to be presented in a subsequent paper (Gregori \etal 1998).
From those calculations
we can anticipate that a magnetic shield always forms and in a fashion 
qualitatively similar to that seen in 2D cases. Moreover, as expected, 
initially elongated clouds develop a stronger magnetic shield than
``spherical'' clouds. However, in the latter case the tension of
the magnetic field lines wrapped around the cloud produces a strong 
deformation of the cloud shape, which grows strongly elongated transverse 
to the plane containing the field and the motion thus facilitating the 
formation of the magnetic shield.

On the other hand the effect of the initial configuration on individual 2D 
cloud evolution has been 
investigated by Miniati \etal (1998) who study the influence of the initial 
field orientation with respect to the cloud motion ($\theta$), of the cloud 
density contrast ($\chi$) and velocity ($M$) on the magnetic bumper formation. 
They find that, as long as 2D approximation is valid and $\theta 
\gtrsim 30^\circ$, the timescale for the 
formation of the magnetic bumper is of the order of 
$\tau \sim (\beta\chi)^{2/3} M^{4/3}~\coll$.
Since clouds are slowed down by the ram pressure of the impinging flux on a 
timescale $\tau_{de}\sim \chi\coll$, they also concluded that magnetic 
bumpers are more likely to develop around high density 
contrast, low Mach number clouds.

Another important 3D issue is the reexpansion of cloud gas
in the direction perpendicular to the computational plane (along the Z-axis).
This effect is important because it could in principle modify our previous
conclusions for the non disruptive cases. However, as it turns out this
only sets a limit on the length of the cloud major axis for the adiabatic Case 6.
In radiative cases lateral reexpansion involves only a small fraction
of cloud mass, independent of the Y or Z direction (see \S \ref{bxsy}).
On the other hand, in the adiabatic case, the rarefaction wave generating such
reexpansion propagates at the sound speed of the postshock
gas, whose temperature has been enhanced by a factor $\sim M^2$. Therefore,
$v_{exp}\sim M c_{si}$. If the length of the cylinder is $\ell_z = \mu R_c$,
the reexpansion occurs on a timescale $\tau_{exp}\simeq \ell_z /v_{exp}
\sim \mu R_c / (M c_{si})=\mu \coll $. Consequently, since
$\rad/\tau_{exp}= \rad / \mu \coll = \eta/\mu $, then if $\mu\gg 1$
($\ell_z \gg R_c$ for cylindrical clouds), the cloud behaves as it did
in 2D radiative cases discussed before.
 
The aforementioned reasons justify 2D simulations as a valuable starting
point for the more complex 3D MHD CCs. As already mentioned along the paper,
however,
in the latter we expect to observe new behaviors not seen in 2D calculations.
We expect those to be mostly related to differences in the evolution of the 
magnetic field. Since the latter is dynamically dominant in transverse
field cases, new behaviors will be more apparent there. For example, even when
the magnetic shield forms in 3D as well, if its strength is much less than in
2D, then we may expect different results from those reported in this paper 
(more precisely in \S \ref{by}). 
Nevertheless,
instabilities and, in general, flows along the third direction, as well as 
quantitative differences in cloud features developed in 2D and 3D,
will certainly affect the dynamics of the collision. 
We have begun a series of 3D MHD cloud simulations to address these
complex issues more fully.

\acknowledgments

FM devotes his efforts in this work to the memory of his friend, 
Leonardo Muzzi, young artist of deep perspective, who inspired his way 
of studying science. We thank R. Dgani for insightful discussions on the
dynamics of magnetized flows,  J. Vall\'ee for useful comments about 
the observed properties of the magnetic field, 
and E. Vazquez Semadeni for stimulating discussions.
The work by FM and TWJ is partially supported by NSF 
grants AST-9318959 and INT-9511654 and
by the University of Minnesota Supercomputer Institute.
The work by DR was supported in part by KOSEF through
grants 975-0200-006-2 and 981-0203-011-2.
AF acknowledges hospitality of University of Minnesota where this work was
started.

\vskip 0.5truecm
\centerline{\bf Appendix A}
\vskip 0.5truecm

We discuss here some details of the treatment of the radiative
losses and of the mass tracing routine.
Radiative losses have been taken into account
using the same approach as in Paper I,
to which we refer for a detailed description.
The radiative correction that we have applied is quasi-second order accurate
(see, \eg LeVeque 1997 for a general discussion).
Let us represent symbolically the ``numerical'' equation as
\begin{equation} 
{\cal D}_t{\bf q} = {\cal D}_s({\bf q}),
\end {equation} 
where ${\cal D}_t$ and ${\cal D}_s$ are the temporal operator and the spatial
plus source operator respectively and ${\bf q}$ is the set of variables
describing our system.
A second order accurate splitting from time step $n$
to $n+2$ for the operator ${\cal D}_s$ would be (Strang 1968):
\begin{equation} 
S_{\frac{1}{2}}L_x L_y S_{\frac{1}{2}}~~S_{\frac{1}{2}}L_y L_x S_{\frac{1}{2}}
\end {equation} 
where $L_x$ and $L_y$ are the differential operator with respect to the \xcoor and 
\ycoor respectively and $S$ is the operator representing
general {\it source} terms. The subscript 
$\frac{1}{2}$ means that the operator is applied for only half of a time step.
Instead, we have used
\begin{equation}
S_{\frac{1}{2}}L_x L_y ~ S~L_y L_x S_{\frac{1}{2}},
\end {equation}
which assumes that $S_{\frac{1}{2}}~~S_{\frac{1}{2}}\equiv S$. Hence the 
``quasi-second'' order description.
In addition we have suppressed cooling inside the shock thickness.
Indeed, in the physical shock layer the flow
should be non-radiative, because the crossing time
of the real shock thickness (artificially spread out by the code)
is much shorter than the cooling time. Also,
since density and pressure are not accurate inside the shock, but only
adjacent to it, radiative cooling
could become artificially large, reducing the performance of the code. This
turns out of particular importance for the MHD calculation.
The radiative cooling function we have used is identical to that 
in Paper I and is fully described in Ferrara \& Field (1994). It 
includes cooling due to free-free emission, recombination lines, 
and collisional excitation lines as well as heating terms provided 
by collisional ionization and ionization by cosmic 
rays. For the cosmic ray ionization rate we adopt the value 
$2\times 10^{-17}$~s$^{-1}$ as determined from observations by
van Dishoeck \& Black (1986). This and other rates can be found in 
Ferrara \& Field (1994) and references therein.
We neglect dust, and particularly
PAHs, photoelectric heating. This is certainly a rough approximation as
long as an accurate model of the multiphase ISM is concerned (\eg Wolfire 
\etal 1995). However, our aim here is to build a simple, albeit reasonable, 
model for the two-phase ISM and concentrate on the properties of collisions 
that do not depend drastically on the details of the multiphase structure.
When pressure equilibrium is imposed, a two-phase (cloud + intercloud)
ISM structure results.

Finally a routine, based on van Leer's second-order
advection scheme (van Leer 1976), has been included to track the fraction of
cloud material inside each grid cell. This quantity, referred to as
``mass tracer'' or ``mass fraction'' (Xu \& Stone, 1995), is initially set 
to unity inside the cloud and zero elsewhere.
The mass fraction allows us to discriminate between the different components
in our simulations, which are the two clouds and the intercloud medium. 
In this way we can calculate various quantities of interest such as
each cloud's kinetic and thermal energy as well as
$Y_{cm}$, the \ycoor of the center of mass of each cloud (Jones \etal 1996).
These are used in the analysis of our results.

\clearpage

\begin{deluxetable}{cccccccc}
\footnotesize
\tablecolumns{8}
\tablecaption{Summary of 2D-MHD Cloud Collisions Simulations \label{tabcs}}
\tablehead{ \colhead{}& \colhead{} & \multicolumn{2}{c}{Clouds Ages} & \colhead{}
 &\colhead{}&\colhead{} &\colhead{} \\
\cline{3-4}\\
\colhead{CASE \tablenotemark{a}} &
\colhead{$\eta$\tablenotemark{b}} &
\colhead{C1} &
\colhead{C2} &
\colhead{$M_r$\tablenotemark{c}} &
\colhead{B} &
\colhead{Grid-Size\tablenotemark{d}} &
\colhead{End Time\tablenotemark{e}}\\
\colhead{}&\colhead{}&\colhead{($\tcr$)}&\colhead{($\tcr$)}&\colhead{}& 
\colhead{} &\colhead{($R_c^2$)}&\colhead{($\coll$)}
}
\startdata
1 & adiabatic & 0.75 & 0.75 & 3 & $B_{x}$ & 15 $\times$ 15 & 75.0 \nl
2 & 0.38      & 0.75 & 0.75 & 3 & $B_{x}$ & 15 $\times$ 10 & 30.0 \nl
3 & adiabatic & 0.5  & 1.0  & 3 & $B_{x}$ & 30 $\times$ 15 & 22.5 \nl
4 & 0.38      & 0.5  & 1.0  & 3 & $B_{x}$ & 30 $\times$ 10 & 22.5 \nl
5 & adiabatic & 0.0  & 0.0  & 3 & $B_{y}$ & 10 $\times$ 20 & 30.0 \nl
6 & adiabatic & 0.75 & 0.75 & 3 & $B_{y}$ & 20 $\times$ 7.5& 30.0 \nl
7 & 0.38      & 0.75 & 0.75 & 3 & $B_{y}$ & 20 $\times$ 7.5& 30.0 \nl
\enddata
\tablenotetext{a}{All models use $\beta=4$, $\gamma$ = 5/3,
$\chi = \rho_c/\rho_i = 100$, equilibrium pressure
$p_{eq}/k_B=1628$~K~cm$^{-3}$. Also, at equilibrium,
we have $T_i= 7400$ K and
$n_i = 0.22~$cm$^{-3}$ for the intercloud medium and
$T_c= 74$ K and $n_c = 22~ $cm$^{-3}$ inside the clouds.}

\tablenotetext{b}{$\eta = \rad/\coll$}

\tablenotetext{c}{This is the {\it relative} Mach number for the cloud pair
when the clouds are first set into motion. It
is referred to the intercloud sound speed, $c_{si}$.}

\tablenotetext{d}{The grid size is expressed in units of cloud radius. One cloud
radius $R_c$=50 zones.}

\tablenotetext{e}{The end time is expressed in terms of collision time $\coll$,
and represents the total time from the beginning of the collision.}
\end{deluxetable}

\begin{deluxetable}{ccccc}
\footnotesize
\tablecolumns{5}
\tablecaption{Clouds Characteristics\tablenotemark{a} \label{tabca}}
\tablehead{\colhead{$\eta$\tablenotemark{b}} &
\colhead{$R_c $} &\colhead{$\coll =R_c/v_c $ } 
&\colhead{$\tcr =2 R_c\sqrt{\chi}/v_c $ }
&\colhead{$\rad =\case{3}{2}\frac{k T}{n\Lambda} $ } \\
\colhead{} & \colhead{(pc)} & \colhead{(yr)} & \colhead{(yr)} & \colhead{(yr)}}

\startdata
Adiabatic  & 0.4 &  2.6$\times 10^4$ &  5.3$\times 10^5$  & $\infty$ \nl
0.38       & 1.5 &  9.7$\times 10^4$ &  2.0$\times 10^6$  & 3.7$\times 10^4$ \nl
\enddata

\tablenotetext{a}{See note {\footnotesize a} in Table \ref{tabcs}.}
\tablenotetext{b}{See note {\footnotesize b} in Table \ref{tabcs}.}
\tablenotetext{c}{The actual value of the cooling time for the smaller cloud is
also $\rad=3.7 \times 10^4$ yr. However, as explained in the text, since 
$\eta > 1$ the cooling does not affect the collision of these clouds and therefore 
it has been turned off during the simulations.
For this reason we have set $\rad=\infty$ in the table.}
\end{deluxetable}

\clearpage

\begin{center}
{\bf FIGURE CAPTIONS}
\end{center}

\figcaption[] {
\label{plot-bx}
Panels showing the evolution of kinetic energy (top left), thermal energy
(bottom left) and \ycm coordinates (bottom right), of the colliding clouds 
and of the total magnetic energy within the grid (top right) as a function 
of time, for the aligned field cases. 
For Cases 1 and 2 the two colliding clouds are identical to each other, 
so only one is displayed. For the first three panels,
solid lines correspond to Case 1, long dash to Case 2. 
Dot and short dash lines respectively 
to cloud C1 (left) and C2 (right) in Case 3. Dot short-dash and
dot-long dash lines respectively to cloud C1 (left) and C2 (right) in Case 4.
For the last panel (top right) solid lines refer to Case 1, dash line to 
Case 2, dot line to Case 3 and dot-long dash line to Case 4.
The kinetic, thermal total magnetic energy in these plots have been normalized
to the sum of the cloud energy (kinetic + thermal + magnetic) and the
background magnetic energy at the beginning of the simulation. Therefore,
with respect to Paper I the vertical scale of the kinetic and thermal 
energy of the cloud is reduced by a factor 1.18.
}

\figcaption[]{ From top to bottom, the four pairs of panels show field line
geometry (left) and density distribution at $t/\coll$ = 4.5, 24, 48.75 and 75.
Coordinates are expressed in units of cloud radii, $R_c$.
Field lines are contours of the magnetic flux, and correspond to change in the
latter by a factor 7. Density figures are inverted grayscale images of 
$f(\rho)=\frac{\rho}{1+\rho}$. 
Top panels show the compression phase $(t=4.5\coll)$ with field lines swept up
by the vertical outflow at the symmetry plane of the collision. 
The following panel shows
the re-expansion phase ($t=24\coll$), with the formation of long finger at the 
back of the cloud, a large spot near the symmetry axis and
a thin shell-like 
structure. Numerous closed field loops witness the occurrence of reconnection 
processes. The next two panels capture the collapse
($t=48.75\coll$) of the large
spot pushed by relaxing field lines. The last two panels 
correspond to the end of the collision ($t=75\coll$) with the formation 
of a thick layer of cloud gas and low magnetic energy. 
\label{bxadsy}}

\figcaption[]{Log density cuts through the grid along the X-axis (y=0.3$R_c$, 
top panel) and along the Y-axis (x=0.1 $R_c$, bottom panel) for Case 1. 
Solid lines refer to $t=4.5 \coll$, dot lines to $t=24\coll$, dash
lines to $t=48.75\coll$ and dot-dash lines to $t=75\coll$.
\label{rho-cut}}

\figcaption[]{Same as Figure \ref{rho-cut}, but for the logarithm 
of thermal pressure. 
\label{pre-cut}}

\figcaption[]{Same as Figure \ref{rho-cut}, but for the logarithm 
of magnetic energy.
\label{bpr-cut}}

\figcaption[] {Field line geometry (left panel) and density distribution
(right panel) for Case 2 at $t=30\coll$. 
Line contours and density images as in Figure \ref{bxadsy}.
\label{bxrasy}}

\figcaption[] {Cuts along the {\it X}-coordinate ($y=0.1R_c$) for the log of 
density (top panel) and log of thermal pressure (bottom panel) for Case 2. 
Solid line corresponds to $t=4.5\coll$, dot line to $t=17.2\coll$ and dash 
line to $t=30\coll$.
\label{cutx-rad}}

\figcaption[]{Field line geometry (top panel) and density distribution
(bottom panel) for Case 3 at $t=22.5\coll$. 
Line contours and density images  as in Figure \ref{bxadsy}.
\label{bxadas}}

\figcaption[] {Field line geometry and density distribution
for Case 4 at $t=6.75\coll$ (a) and $t=22.5\coll$ 
(b) respectively. Line contours and density images are 
created as in Figure \ref{bxadsy}.
\label{bxraas}}

\figcaption[] {Density distribution with superimposed field line geometry 
for Case 5 at $t=7.5\coll$ (left panel) and $t=30\coll$ (right panel).
Line contours and density images are created as in Figure \ref{bxadsy}.
\label{byad0}}

\figcaption[] {Plots as in Figure \ref{plot-bx} but for transverse field cases.
Dash lines refer to Case 5, solid and dot lines to Cases 6 and 7 
respectively. The latter two lines have been multiplied by a factor 10 and 100 
respectively, in order to make them readable on the same vertical scale
as the dashed line. 
\label{plot-by}}

\figcaption[] {Field line geometry and density distribution
for Case 6 at $t=12\coll$ (a) and $t=30\coll$ 
(b) respectively. Note that, since the magnetic field is quite stronger 
than in previous cases, contours of the 
magnetic flux (for the field lines) correspond to change in the
latter by a factor 14 (instead of 7 as before). 
Density images, on the other hand, are as in Figure \ref{bxadsy}.
\label{byadsy}}

\figcaption[] {Field line geometry and density distribution
for Case 7 at $t=9.75\coll$ (a) and $t=22.5\coll$ 
(b) respectively. Line contours and density images 
as in Figure \ref{byadsy}.
\label{byrasy}}

\end{document}